# All Electronic Isolator Action Device Based on Negative Refractive Heterostructure Bi-Crystal with Field Induced Asymmetry

Clifford M. Krowne

Microwave Technology Branch, Electronics Science & Technology Division, Naval Research Laboratory, Washington, DC 20375

## Abstract

It has been discovered that heterostructure bi-crystal arrangements lead to field asymmetry in guided wave structures. Here a study is conducted over a range of nominal permittivity values to see if the effect is present in widely varying dielectric materials. Marked shifts of the field distribution occurs in some cases, and this can be the basis of an all electronic isolator. Such an all electronic device could be fixed or even constructed as a control component using materials with electrostatically controllable permittivity. Distributions have been obtained to demonstrate the effect using an anisotropic Green's function solver.

# I. Introduction

Recently it has been shown in the physics community that a bi-crystalline pair of materials leads to field asymmetry [1]. This was accomplished by realizing that a similar arrangement consisting of two crystals, properly oriented with respect to each other, provides a structure capable of producing negative refraction for some directions of the incoming wave in an optical scattering numerical and experimental test [2]. Field asymmetry arises from some properties of the broken symmetry, only available by using a heterostructure. The simplest arrangement is the bi-crystal pair, with the crystals chosen as uniaxial, possessing two ordinary principal axes, and one extraordinary principal axis. Although the original discovery of field asymmetric was for nominal values of permittivity $\varepsilon = 5$, there is no reason why the effect can not be found in other crystalline materials. It is only necessary to utilize the uniaxial properties of the crystal.

In fact, it is this universality of the effect, which leads us to the next conclusion, that it is possible to produce the effect starting with isotropic crystals if they are ferroelectric. Logic is as follows. Start with a ferroelectric crystal which is isotropic. Apply a static electric bias field $E_0$ is some direction, and increase the field until the desired reduction in permittivity occurs in the bias direction. The artificially induced preferred direction, becomes the extraordinary direction and is a principal axis direction. Permittivity tensor element in that direction is the extraordinary permittivity diagonal value $\varepsilon_e$. Two other principal axis directions, normal to this preferred biased direction, become the ordinary directions and in those directions is the unbiased original permittivity, equal to the ordinary permittivity $\varepsilon_o$.

Ferroelectric behavior of permittivity change is based upon a phase transition, going from a cubic to tetragonal atomic crystalline arrangement, which takes the crystal from a paraelectric state to a ferreoelectric state. This is why ferroelectric materials are so attractive for electronic applications, because huge percentage changes in the dielectric constant may be made. For example, it is typical to be able to obtain, at proper temperatures, nominal values of permittivity being 2000, 500, or 140, and being able to get final values of 400, 250, and 100, These dielectric constant final values correspond to 5:1, 2:1 and to 29 % changes using suitable bias $E_0$ values.

So two possibilities exist. The first one is simply to acquire uniaxial crystals, properly orient their crystalline planes (to be covered in detail in the next section), and build the structure to provide a fixed given asymmetry. Second possibility, is to work with a ferroelectric crystalline system, and implement biasing configurations dc isolated from the rf characteristics of the electromagnetic structure, allowing variable asymmetry. First structure realized does not need any external static electric field biasing. Second structure realized requires biasing configurations, and is more complicated, but has the tremendous quality of being a variable control component. The second structure, creates what is termed a negative crystal [3], because the extraordinary permittivity value is deflated compared to the ordinary permittivity value.

## II. Theoretical Crystal Tensor Rotations

Bi-crystal layering which produces the effect has two adjacent layers with opposite rotations of the principal cross-sectional axes, the rotation angles denoted by $\theta$, where the positive angle corresponds to a counter-clockwise rotation of the cross-sectional xy axes about the z-axis. Electromagnetic waves propagate down the z-axis, the longitudinal axis of the uniform guiding structure. To utilize the negative refractive property, the guiding metal is placed between the two crystals. In such an arrangement, one crystal is the bottom substrate, the other crystal the superstrate on top. The tensor of an unrotated, principal axis system crystal, is given by the dyadic $\bar{\bar{\varepsilon}} = \varepsilon_{xx}\hat{x}\hat{x} + \varepsilon_{yy}\hat{y}\hat{y} + \varepsilon_{zz}\hat{z}\hat{z}$. Rotation of the bottom crystal about the z-axis in the $\theta = \theta_b$ amount creates off diagonal elements equal to $\varepsilon_{rxy} = \varepsilon_{ryx} = (\varepsilon_{yy} - \varepsilon_{xx})\sin 2\theta_b/2$ with the last dyadic term unchanged [1], [4]. All other off-diagonal elements remain zero. Diagonal elements become $\varepsilon_{rxx} = (\varepsilon_{xx}\cos^2\theta_b + \varepsilon_{yy}\sin^2\theta_b)$ and $\varepsilon_{ryy} = (\varepsilon_{xx}\sin^2\theta_b + \varepsilon_{yy}\cos^2\theta_b)$. Rotation of the top crystal about the z-axis in the $\theta = \theta_t$ amount creates off-diagonal elements equal to $\varepsilon_{rxy} = \varepsilon_{ryx} = (\varepsilon_{yy} - \varepsilon_{xx})\sin 2\theta_t/2$ with the last dyadic term unchanged. All other off-diagonal elements remain zero. Diagonal elements become $\varepsilon_{rxx} = (\varepsilon_{xx}\cos^2\theta_t + \varepsilon_{yy}\sin^2\theta_t)$ and $\varepsilon_{ryy} = (\varepsilon_{xx}\sin^2\theta_t + \varepsilon_{yy}\cos^2\theta_t)$.

One notices that the off-diagonal elements are maximized in the individual layers when $\sin 2\theta = 1$. This occurs for $\theta = \pm 45$. Increasing the angle still further, up to $\theta = \pm 90$, however, supposing $\theta_b = 90$ and $\theta_t = -90$, causes the incommensurate marking off of atomic layers to vanish, making symmetry appear again (looks like the unrotated case again). Generally one expects $\theta_b - \theta_t = \pm 90$ to produce the maximum effect.

There are only two possibilities for picking out the beginning unrotated principal axis system uniaxial tenors. Dyadics must be either $\bar{\bar{\varepsilon}} = \varepsilon_e\hat{x}\hat{x} + \varepsilon_o\hat{y}\hat{y} + \varepsilon_o\hat{z}\hat{z}$ or $\bar{\bar{\varepsilon}} = \varepsilon_o\hat{x}\hat{x} + \varepsilon_e\hat{y}\hat{y} + \varepsilon_o\hat{z}\hat{z}$ because there are only two ways to insert the single extraordinary axis permittivity into the $2 \times 2$ submatrix, also forcing the last diagonal element to be the ordinary value. We choose the first case.

If $\theta = \pm 45$ is selected, with $\theta_b = 45$ and $\theta_t = -45$, $\sin^2\theta = \cos^2\theta$ making diagonal elements $\varepsilon_{rxx} = \varepsilon_{ryy} = (\varepsilon_{xx} + \varepsilon_{yy})/2 = \varepsilon_a$, an averaged value of the first two principal axis diagonal elements. It must be $\varepsilon_a = (\varepsilon_e + \varepsilon_o)/2$. Off-diagonal elements are $\varepsilon_{rxy} = \varepsilon_{ryx} = \pm (\varepsilon_{yy} - \varepsilon_{xx})/2 = \pm \varepsilon_d$ with $\theta > 0$ or $\theta < 0$ for the plus or minus signs on $\varepsilon_d$. So $\varepsilon_d = (\varepsilon_o - \varepsilon_e)/2$ for a negative crystal.

## III. Guided Stripline Structure

Structure to be studied numerically here is a single stripline configuration with bounding vertical walls and a ground plane and a top cover. Although results will be obtained for the symmetric geometric placement of the strip with respect to all the bounding walls, better to unambiguously show that any asymmetry of the fields must come from the crystalline properties of the bi - layer arrangement, there is no reason in principle why, for example, each crystal layer can not be of unequal thickness, causing the field to be unsymmetric in the vertical direction. Figure 1 shows a cross-sectional drawing of the structure. For calculations to be done in the next section, we take $w = h_T = h_B = 5$ mm, $b = 50$ mm, making $h_{TOTAL} = h_T + h_B = 10$ mm. $b/w = 10$.

Cross-hatching in Fig. 1 is meant to show the crystalline planes, and the normal to them indicates a principal axis direction for each one of the crystals. Strip thickness is taken to be vanishing small.

## IV. Electromagnetic Fields

Starting with the structure in Fig. 1, computations were run for nominal values of the permittivity $\varepsilon = 500$, 140, and 30. Fig. 2 shows the electromagnetic field distribution for the transverse electric field $\mathbf{E}_t$ in the cross-section. $\mathbf{H}_t$ is similar and won't be shown here due to space limitations. Frequency was $f = 10$ GHz and the propagation constant pure phase with $\gamma = \alpha + j\beta = j\beta$, $\beta = 4.392$ normalized to the free space value. Number of even and odd current basis functions was $n_x = n_z = 1$ for currents in the x and z directions. Both parities of the basis functions is needed to allow for asymmetric current distributions in the x direction. Number of spectral terms was $n = 200$. Permittivity values were $\varepsilon_e = 15$ and $\varepsilon_o = 30$, making $\varepsilon_a = 22.5$ and $\varepsilon_d = 7.5$. $\varepsilon_d/\varepsilon_a = 33$ %. Distribution is Fig. 2 seems to be a fundamental mode fixed about the strip, with a cycloid shape, and the major intensity of the distribution shifted to the left. (Strip located at $|x| \leq 2.5$ mm or $-2.5 \leq x \leq 2.5$.)

Fig. 3 shows the transverse electric field $\mathbf{E}_t$ for $f = 10$ GHz. Permittivity values were $\varepsilon_e = 110$ and $\varepsilon_o = 140$, making $\varepsilon_a = 125$ and $\varepsilon_d = 15$. $\varepsilon_d/\varepsilon_a = 12$ %. Distribution in Fig. 3 also seems to be a fundamental mode but with less of a pronounced cycloid shape than before. Overall intensity of the entire distribution is even more shifted to the left. Again the propagation constant is pure phase with $\beta = 10.38$ normalized to the free space value.

Lastly, Fig. 4 shows the transverse electric field $\mathbf{E}_t$ for $f = 2$ GHz. Permittivity values were $\varepsilon_e = 250$ and $\varepsilon_o = 500$, making $\varepsilon_a = 375$ and $\varepsilon_d = 125$. $\varepsilon_d/\varepsilon_a = 33$ %. Distribution in Fig. 4 seems to be a fundamental mode, or at least one that is close to being the fundamental, as the majority of the distribution's highest strength is located about the strip. Elliptical distribution shapes appear. Marked shift of the overall distribution to the left is apparent. Again the propagation constant is pure phase with $\beta = 17.888$ normalized to the free space value.

Note that all the calculations were performed using an anisotropic Green's function spectral domain method [5], [6].

## V. Surface Current Distribution

As one would suspect, the surface current distributions for each of the cases just examined for the electric field distributions in Figs. 2 – 4 are asymmetric. Fig. 5 provides the surface current corresponding to Fig 4.

## VI. Isolator Action

Isolator action can be enabled by taking advantage of the asymmetric electromagnetic field distribution by inserting a lossy strip, a second strip, beside the symmetrically located guiding strip, so that it is off – centered and positioned correctly so as to attenuate the wave when the direction is reversed from the low loss direction. This concept is well known, and is referred to as the field displacement effect, and has been widely employed in nonreciprocally based isolation devices, often utilizing ferrite material.

If the field displacement effect is employed in the bi–crystal heterostructure, a device can be built without the need of magnets. Even for the bi–crystal heterostructure with tuning capability based upon ferroelectric materials [7], [8], [9], only electric fields are used to bias the device. (See Fig. 1, top crystal, which shows biasing dc circuit.) A special advantage may accrue to using ferroelectric materials, in that even for the situation where one has amorphous material with random micro – crystal orientations, imposition of a biasing field may allow artificial creation of the principal axes, a requirement for getting the bi–crystal to exist.

It may be desirable to actively sense whether the wave enters from port 1 (into the page – see Fig. 1) or port 2 (out of the page) and electronically bias the ferroelectric crystals to shift the rf field magnitude to be low loss or high loss with regard to the lossy strip.

# VII. Conclusion

It has been demonstrated here that a new arrangement of crystals into heterostructures can produce asymmetric field distributions, with the potential possibly of leading to new and useful isolator devices at microwave frequencies.

Figure Captions

1. Cross-section of the bi-crystal structure. Differently oriented crystals sandwich the strip. Biasing circuit shown for the upper half of the structure – lower half is similar.
2. Electric field distribution for nominal $\varepsilon = 30$. Frequency $f = 10$ GHz.
3. Electric field distribution for nominal $\varepsilon = 140$. Frequency $f = 10$ GHz.
4. Electric field distribution for nominal $\varepsilon = 500$. Frequency $f = 2$ GHz.
5. Surface current distribution corresponding to Fig. 4.

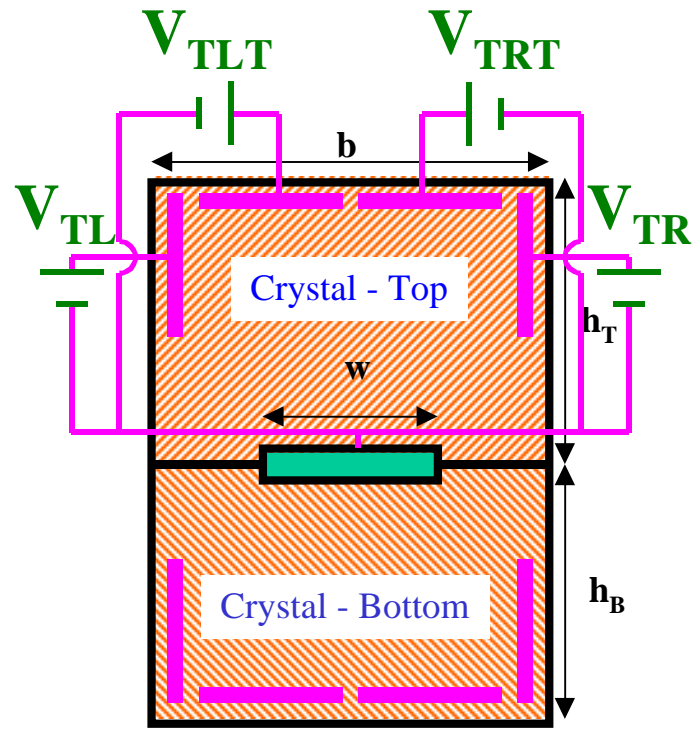

FIGURE 1

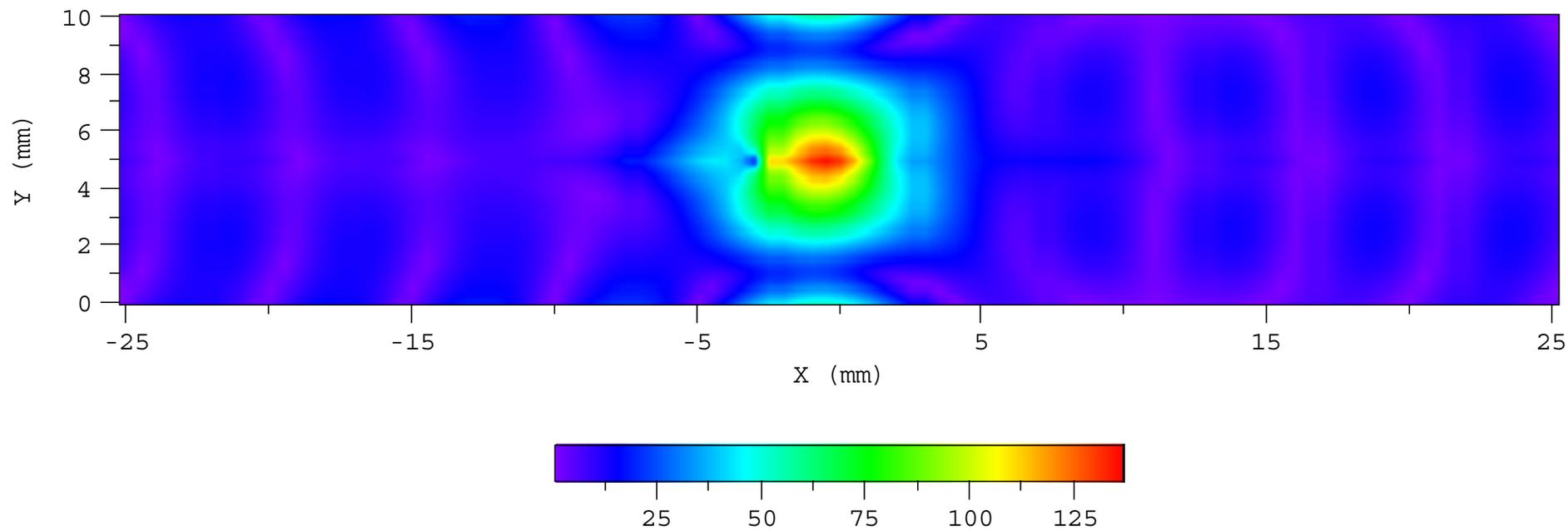

FIGURE 2

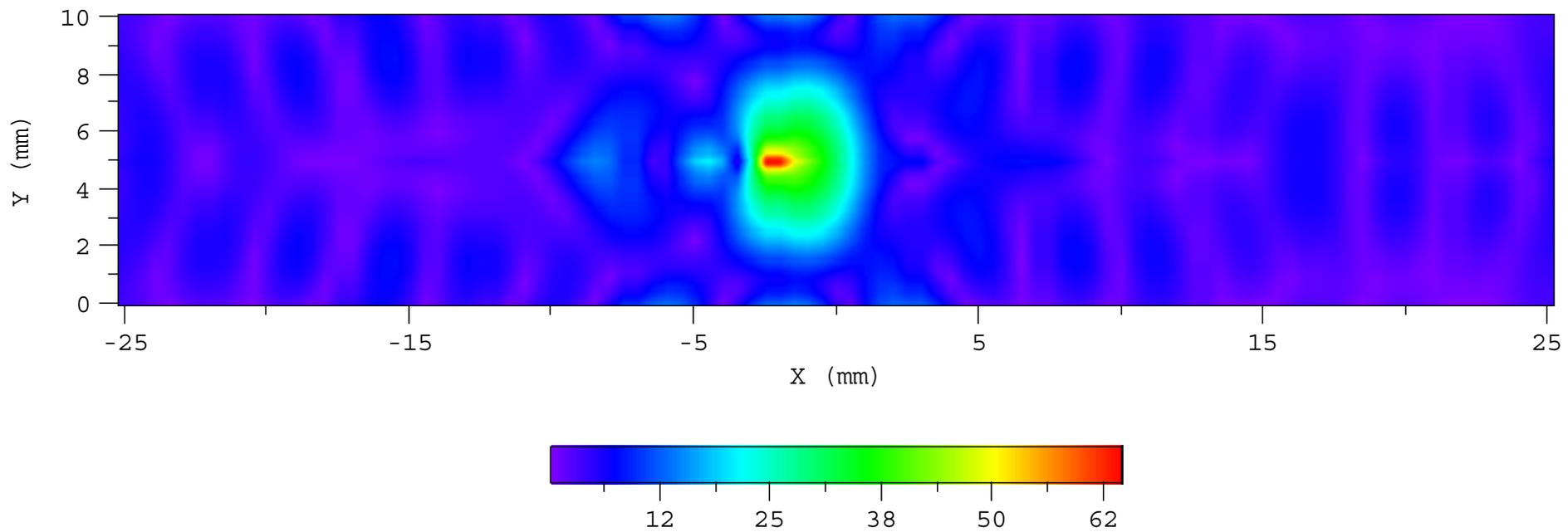

FIGURE 3

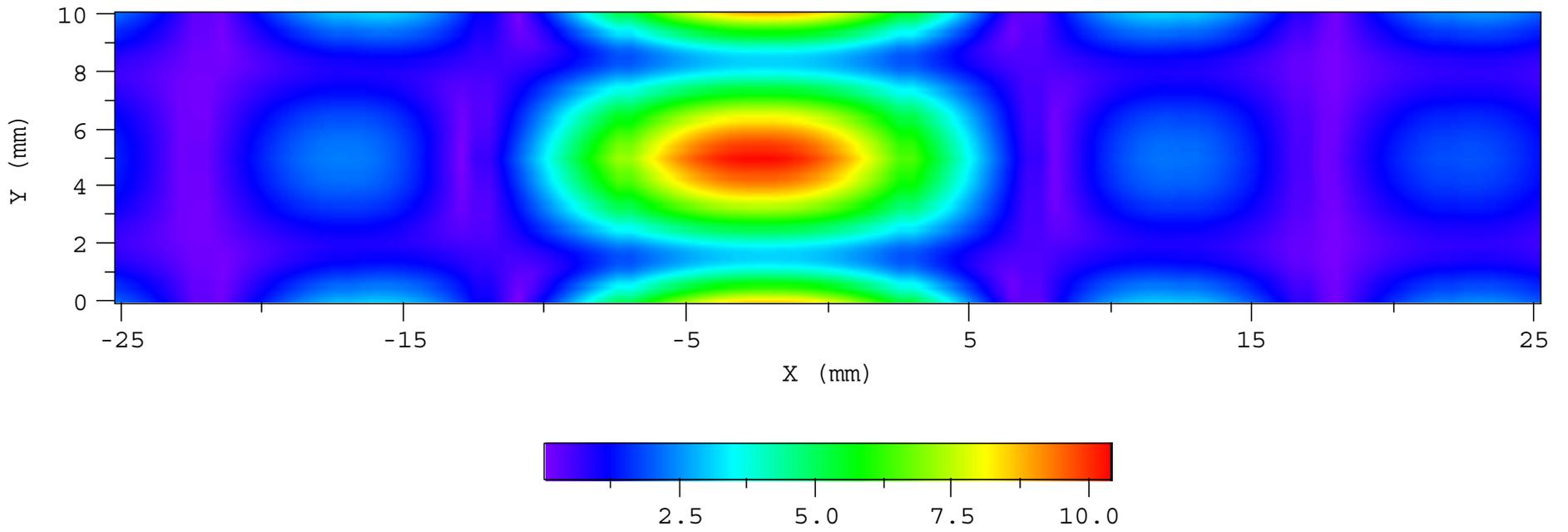

FIGURE 4

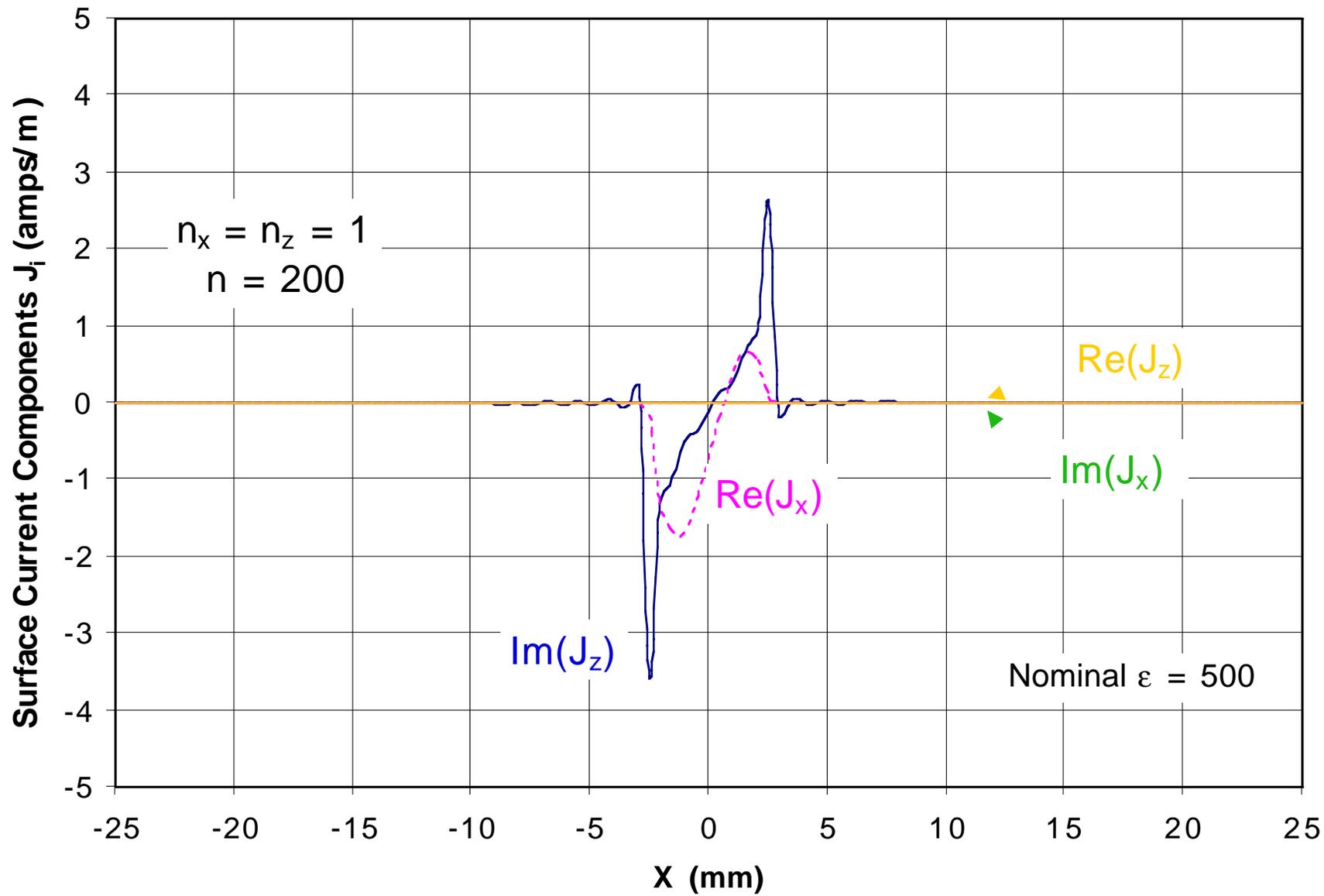

FIGURE 5